\documentclass[journal]{IEEEtran}
\usepackage[export]{adjustbox}  
\usepackage{amssymb, amsmath}
\usepackage{color,soul} 

\usepackage{jabbrv}
\usepackage{cite}

\ifCLASSINFOpdf
\else
\fi

\begin{document}
%
\title{Solving multi-armed bandit problems using a chaotic microresonator comb}
%
%
%

\author{Jonathan Cuevas$^1$, Ryugo Iwami$^2$, Atsushi Uchida$^2$, 
        Kaoru Minoshima$^{3,4}$,
        and Naoya Kuse$^{4,*}$

\thanks{$^1$Graduate School of Sciences and Technology for Innovation, Tokushima University, 2-1, Minami-Josanjima, Tokushima, Tokushima 770-8506, Japan, $^2$Department of Information and Computer Sciences, Saitama University, 255 Shimo-okubo, Sakura-ku, Saitama City, Saitama 338-8570, Japan, $^3$Graduate School of Informatics and Engineering, The University of Electro-Communications, 1-5-1 Chofugaoka, Chofu, Tokyo 182-8585, Japan, $^4$Institute of Post-LED Photonics, Tokushima University, 2-1, Minami-Josanjima, Tokushima, Tokushima 770-8506, Japan. \\E-mail: kuse.naoya@tokushima-u.ac.jp}}

%
%

\markboth{Solving multi-armed bandit problems using a chaotic microresonator comb}%
{Shell \MakeLowercase{\textit{et al.}}: Bare Demo of IEEEtran.cls for IEEE Journals}
%



\maketitle

\begin{abstract}
The Multi-Armed Bandit (MAB) problem, foundational to reinforcement learning-based decision-making, addresses the challenge of maximizing rewards amidst multiple uncertain choices. While algorithmic solutions are effective, their computational efficiency diminishes with increasing problem complexity. Photonic accelerators, leveraging temporal and spatial-temporal chaos, have emerged as promising alternatives. However, despite these advancements, current approaches either compromise computation speed or amplify system complexity. In this paper, we introduce a chaotic microresonator frequency comb (chaos comb) to tackle the MAB problem, where each comb mode is assigned to a slot machine. Through a proof-of-concept experiment, we employ 44 comb modes to address an MAB with 44 slot machines, demonstrating performance competitive with both conventional software algorithms and other photonic methods. Further, the scalability of decision making is explored with up to 512 slot machines using experimentally obtained temporal chaos in different time slots. Power-law scalability is achieved with an exponent of 0.96, outperforming conventional software-based algorithms. Moreover, we find that a numerically calculated chaos comb accurately reproduces experimental results, paving the way for discussions on strategies to increase the number of slot machines.
\end{abstract}

%
\IEEEpeerreviewmaketitle

\section{Introduction}
The Multi-Armed Bandit (MAB) problem has become a cornerstone of decision-making processes based on reinforcement learning  \cite{Slivkins2019}. Its name, suggesting a gambler faced with many slot machines (or "bandits"), captures its central challenge: an agent's pursuit to maximize rewards while navigating the uncertainty of multiple choices. This problem reflects real-world situations that range from customizing online advertisements for individual users to optimizing clinical trials in the medical field. Central to the MAB problem is a delicate balance—an agent must not only strive to accrue maximum rewards but must also invest in exploration to identify the most rewarding option from a set. 

Traditional algorithmic approaches, while effective, often encounter computational challenges as the complexity (i.e., the number of slot machines, $N_{\rm slot}$) of the problem grows. This has led to innovative solutions that utilize the computational power of photonics, referred to as "photonic accelerators" in this research field \cite{Shastri2021,Kitayama2019}. Initial experiments employed temporal chaos generated from a self-injection locked (SIL) semiconductor laser \cite{Sciamanna2015}. By setting a threshold to this temporal chaos, two-armed bandit problems were addressed \cite{Naruse2017}. Through the use of time-division multiplexing, $N_{\rm slot}$ has been expanded to 64 \cite{Naruse2018}. However, time-division multiplexing reduces computation speed. To address the trade-off between $N_{\rm slot}$ and computation speed, other photonic methods have been explored \cite{Mihana2019, Mihana2020,Mihana2019, Mihana2020,morijiri2022decision,iwami2022controlling,morijiri2023parallel}. Using coupled \cite{Mihana2019, Mihana2020} or independent semiconductor lasers \cite{morijiri2022decision} increases $N_{\rm slot}$ up to the number of lasers. In this setup, each laser corresponds to a slot machine. However, this comes with increased system complexity, requiring a laser for each slot.
Another strategy employs a multimode laser, where each longitudinal mode of the laser corresponds to a slot machine \cite{iwami2022controlling}. By injecting single-mode lasers with carefully selected frequencies and powers, it becomes possible to target a specific longitudinal mode, effectively choosing a slot machine with the highest reward. While numerical models show promising scaling in terms of the correct decision rate (CDR) over $N_{\rm play}$ (the likelihood to select the highest-reward slot machine on the $N_{\rm play}$ th try), practical implementations face challenges. The system's scalability is limited by the need for many single-frequency lasers, each set to a specific frequency. There's also a limit to the number of longitudinal modes a multimode laser can support.
A different method involves optical spatiotemporal chaos, which uses a spatial light modulator (SLM) and a CMOS camera \cite{morijiri2023parallel}. Here, each of the SLM and camera corresponds to a slot machine. This can potentially handle a large number of machines. However, this method has limitations. The speed at which chaos is generated is bound by the SLM and CMOS camera's capabilities, often not reaching the desired kHz range.

Chaotic microresonator combs, also known as "chaos combs," are generated by feeding a single-frequency laser into a high-Q microresonator \cite{Herr2012} through cascaded four-wave mixing (FWM) within the microresonator. Since both components can be fabricated using a CMOS-compatible process, chaos combs are suitable for mass production and can be implemented on a chip scale \cite{xiang2021laser}. Due to the highly nonlinear and infinite-dimensional systems present in high-Q microresonators \cite{Godey2014}, chaos combs can form dozens of chaotic comb modes that exhibit no correlation with one another. Leveraging this lack of correlation, chaos combs have been utilized to parallelize chaos LiDAR, enabling faster acquisition speeds \cite{chen2023breaking,lukashchuk2023chaotic1,lukashchuk2023chaotic2}.

In this paper, we provide comprehensive investigation on the application of a chaos comb to address the MAB problem. This approach ensures that neither system scalability is compromised nor the generation rate of chaos diminished. In our method, the comb modes of the chaos comb are allocated to slot machines. In a proof-of-concept experiment, 44 comb modes are employed, allowing tackling of an MAB with $N_{\rm slot}$ = 44. By leveraging multiple (up to 512) temporal chaos signals from various comb modes across different time slots, we precisely scale the CDR to achieve 0.95 ($N_{\rm CDR095}$) with a predicted relationship of $17.4 \times N_{\rm slot}{}^{0.96}$. This outcome outperforms traditional software algorithms, such as UCB1-tuned and Thompson sampling, and surpasses other photonic methods. Furthermore, a  numerically calculated chaos comb based on our experimental parameters is presented. This reveals a strong match between the chaos features of the comb, obtained both experimentally and numerically (including the optical/RF spectrum and the autocorrelation (ACF) of the temporal chaos), and the results from the MAB experiments, such as CDR, regret, and entropy. This insight provides strategies to increase $N_{\rm slot}$ using chaos combs.


\section{Concept and experimental setup}
Figure 1(a) shows a conceptual illustration of the proposed method. The comb modes of a chaos comb are separated using a wavelength-division multiplexer, yielding multiple temporal chaos outputs assigned to slot machines. $N_{\rm slot}$ matches the number of comb modes. Given that a chaos comb has an optical spectrum spanning over 50 nm with a comb mode spacing of 100 GHz at telecom wavelengths, we can easily manage a few dozen slot machines.

The experimental setup for our proof-of-concept is illustrated in Fig. 1(b). We employ a single-frequency CW laser oscillating around 1553 nm as our pump laser. This CW laser is amplified by an erbium-doped fiber amplifier (EDFA) to 300 mW. The amplified light is then coupled into a Si$_3$N$_4$ microresonator using a lensed fiber. The light's polarization is adjusted using a polarization controller (not shown in Fig. 1(b)). The Si$_3$N$_4$ microresonator \cite{Pfeiffer2018,Liu2021}, has a loaded Q factor of around 95 MHz and a free-spectral range (FSR) of approximately 95 GHz. A chaos comb is produced by adjusting the frequency of the pump laser from blue to red. The detuning in frequency between the pump laser and microresonator is set just before the chaos comb transitions to a dissipative Kerr soliton comb \cite{Kippenberg2018}. The residual pump laser on top of the chaos comb is removed using a bandstop filter. This is followed by an optical waveshaper, which extracts four comb modes. The output from the waveshaper is then amplified by another EDFA, providing an optical power of about 20-30 $\mu$W for each comb mode. These are then sent to photodetectors in the subsequent setup.
The WDM, which separates the four comb modes, is composed of three 1 by 2 optical couplers (with a 50:50 splitting ratio) combined with four optical bandpass filters. Each filter extracts only one of the four comb modes based on its mode number. Finally, the separated comb modes are sent to photodetectors (PDB480C-AC with a bandwidth of 1.6 GHz from Thorlabs) and are digitized using an oscilloscope (MSO8204 with a bandwidth and used sampling rate of 2 GHz and 2.5 GSa/s, respectively, from RIGOL).

\begin{figure}[!h]
\centering
\fbox{\includegraphics[width=0.95\linewidth]{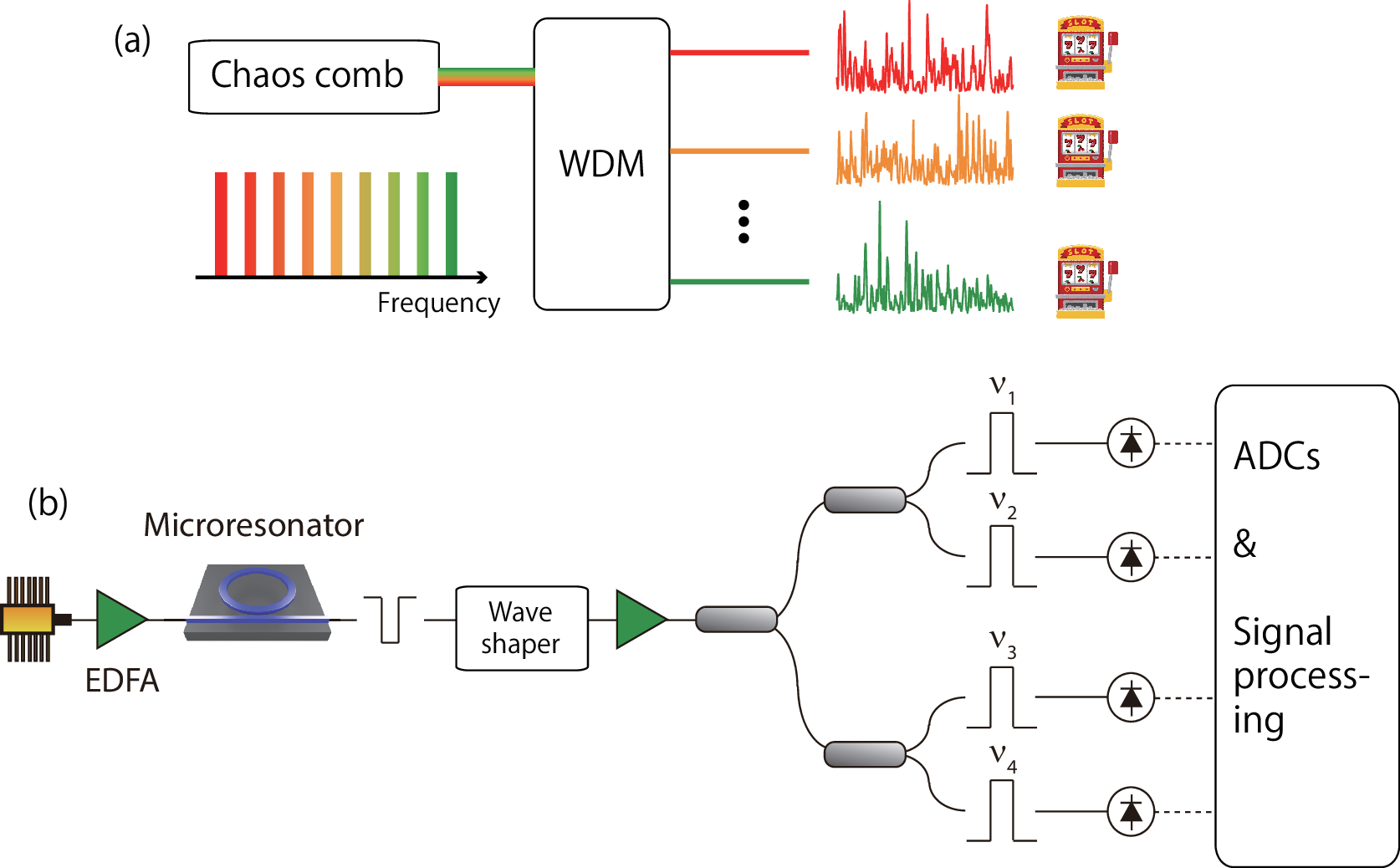}}
\caption{(a) Conceptual schematic of the proposed method. Comb modes of a chaos comb is allocated to slot machines. (b) Schematic of the experimental setup. $\nu_i$ (i = 1, 2, 3, 4) stands for optical frequencies. WDM: wavelength-division multiplexer, EDFA: erbium-doped fiber amplifier, ADC: analog-to-digital converter.}
\end{figure}

\section{Characterization of chaos comb}
Before addressing the MAB problem, the chaos comb is characterized both experimentally and numerically. In the experiment, we measure the optical spectra and chaos signals in both time and frequency domains. As depicted in Fig. 2(a), the optical spectrum has a 10-dB bandwidth of roughly 12 THz (around 100 nm), offering 130 comb modes within this bandwidth. An example of temporal chaos from a single comb mode is presented in Fig. 2(b). The chaos's immediate response time is about 1 ns, which can be further reduced by using a higher pump power and a lower loaded Q combined with a higher nonlinear index \cite{Chang2020}. The temporal chaos exhibits an RF 10-dB bandwidth of approximately 750 MHz, as displayed in Fig. 2(c) (detected by a PD (RX10AF with a bandwidth of 10 GHz from Thorlabs)), with minor peaks at 280 and 560 MHz. These peaks result in small spikes in the ACF around 3.5 ns (as shown in Fig. 2(d)). The ACF has a full width at half maximum (FWHM) of 1.5 ns. The numerically simulated chaos comb is then compared to the experimentally observed chaos comb. For numerical simulation, we used the normalized Lugiato-Lefever Equation (LLE) \cite{Coen2012,Chembo2013} to calculate the intraresonator field ($\psi$):
\begin{equation}
    \frac{\partial}{\partial t}\psi = -(1+i\delta)\psi + i|\psi|^2\psi + i\frac{D_2}{2!} \left(i\frac{\partial }{\partial \theta }\right)^2\psi +F.
\end{equation}

In this equation, $t$ is the slow time representing the evolution of the intraresonator field over many round trips, and $\theta$ is a fast-time variable corresponding to the intraresonator angle in a moving reference frame. $\delta$, $D_2$, and $F^2$ are the normalized detuning, second-order dispersion, and pump power, respectively. Dispersion is adjusted to match the comb mode number for the initial FWM between the calculation and experiment. $F^2$ of 20 is chosen based on the experimentally measured ratio between the utilized pump power and the threshold pump power required to initiate the first four-wave mixing (FWM). The detuning is adjusted to the value just before the transition to a breather or soliton comb. An FSR of 95 GHz and a loaded Q of 95 MHz are applied to give the chaos combs a physical time and frequency dimension.

\begin{figure}[!b]
\centering
\fbox{\includegraphics[width=0.95\linewidth]{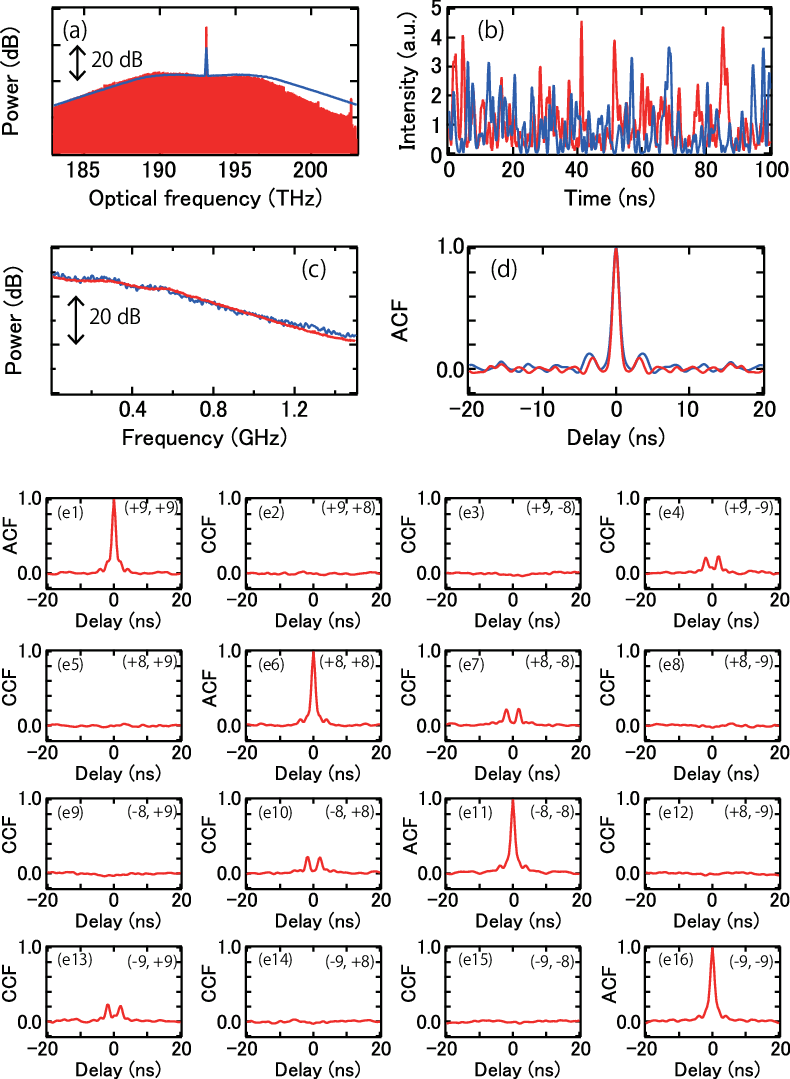}}
\caption{(a) Optical spectra of the chaos comb obtained from the experiment (red) and numerical calculation (blue). (b) Temporal chaos generated from the experimentally (red) and numerically (blue) obtained chaos comb. Sampling rate is 10 GSa/s. (c) RF spectra of the chaos comb obtained from the experiment (red) and numerical calculation (blue). (d) Autocorrelation (ACF) of the temporal chaos from the experiment (red) and numerical calculation (blue) obtained from the temporal chaos shown in (b). (e) ACFs and crosscorrelations (CCFs) between the two of (+9, +8, -8, -9) th comb modes.}
\end{figure}

Although the experimental optical spectrum displays asymmetry, the calculated spectrum is symmetrical. This discrepancy arises because the calculation does not account for frequency-dependent output coupling to the straight waveguide \cite{Moille2019}, as well as higher-order ($\geq$ 3) dispersion \cite{Bao2017} and the Raman effect \cite{Karpov2016}. Nevertheless, the temporal chaos from the calculation closely mirrors the experimental results, as depicted in Fig. 2(a). The numerically calculated RF spectrum and ACF are compared to experimental findings, as shown in Figs. 2(c) and (d). The calculated RF spectrum differs by only a few dB from the experimental results across a broad range of offset frequencies. The FWHM of the ACF is also consistent within 10 \%. These findings confirm that the numerical calculations closely match the experimentally derived chaos comb. Therefore, numerically obtained chaos combs are employed to explore the scaling of slot machines in the MAB problem.

In addition to ACFs, we also calculated cross-correlations (CCFs) between the temporal chaos from different comb modes. Figure 2(e) displays ACFs and CCFs for the -9th, -8th, 8th, and 9th comb modes. As anticipated from the generation mechanism (i.e., FWM) of the chaos comb, pairs (-9th and 9th) and (-8th and 8th) exhibit correlation. This is because these pairs of photons are generated via FWM simultaneously. The reason for the split in the CCFs remains unclear. In our numerical calculations, this split isn't observed, and a single peak appears. Aside from these pairs, no correlation is detected, indicating that half of the chaos comb's modes can be viewed as independent temporal chaos. 


\section{Solving MAB}
For the MAB experiment, we use comb modes ranging from the -19th to the +35th (excluding -3rd to +7th due to optical filtering), totaling 44 modes. Due to equipment limitations, only four comb modes can be detected simultaneously. Temporal chaos from other comb modes is collected in separate time slots. For this proof-of-concept experiment, we use both + and - order comb modes. However, note that if all comb modes are used simultaneously, the cross-correlation between pairs of +/- comb modes might influence the results of MAB. The oscilloscope's sampling rate for this part is 2.5 GSa/s. After digitizing data from the 44 temporal chaos, the MAB problem is addressed through offline signal processing. This offline processing encompasses the normalization of the temporal chaos and the selection of the played slot according to the following algorithm. The hit probability is set as 0.9, 0.7, 0.5, 0.1 for the initial four slot machines and alternates between 0.7 and 0.5 for the subsequent machines, following the model in reference \cite{morijiri2022decision,morijiri2023parallel} for fair comparison. The selection of played slot machines is based on the tug-of-war algorithm \cite{Kim2010,morijiri2022decision}. In this algorithm, the slot machine with the largest biased-chaos signal, which is the sum of a chaos signal and a bias $B_m(N_{\rm play})$ (where $m$ is the label of the slot machine and $N_{\rm play}$ represents the number of plays) is chosen. $B_m(N_{\rm play})$ is calculated as:
\begin{equation}
    B_m(N_{\rm play}) = Q_m(N_{\rm play}) - \frac{1}{N_{\rm slot}-1}\sum_{m'\neq m}^{N_{\rm slot}}Q_{m'}(N_{\rm play}),
\end{equation}
\begin{equation}
    Q_m(N_{\rm play}) = N_{\mathrm{play},m} - \left(1 + \frac{\hat{P}_{\rm top1} + \hat{P}_{\rm top2}}{2 - \hat{P}_{\rm top1} - \hat{P}_{\rm top2}}\right) N_{\mathrm{loss},m}.
\end{equation}
Here, $Q_m(N_{\rm play})$ is termed the evaluation value, while $N_{\mathrm{play},m}$, and $N_{\mathrm{loss},m}$ represent the number of plays and losses for the $m$th slot machine, respectively. $\hat{P}_{\rm top1}$ and $\hat{P}_{\rm top2}$ are the first and second highest estimated hit probabilities, which are derived by calculating $\frac{N_{\mathrm{play},m} - N_{\mathrm{loss},m}}{N_{\mathrm{play},m}}$

An example of the temporal chaos with biases determined by the tug-of-war algorithm is presented in Fig. 3(a) when $N_{\rm slot}$ is 8. Initially, the selection of slot machines appears somewhat random, signifying exploration. After a few plays, the signal from the first slot machine grows stronger, reducing the likelihood of selecting less rewarding slot machines, leading to exploitation. Figure 3(b) displays the selected slot machines at the $N$th play when $N_{\rm slot}$ is 44, providing a clearer view of the exploration and exploitation phases. Exploration concludes around the 400th play, and exploitation begins.

\begin{figure}[!ht]
\centering
\fbox{\includegraphics[width=0.95\linewidth]{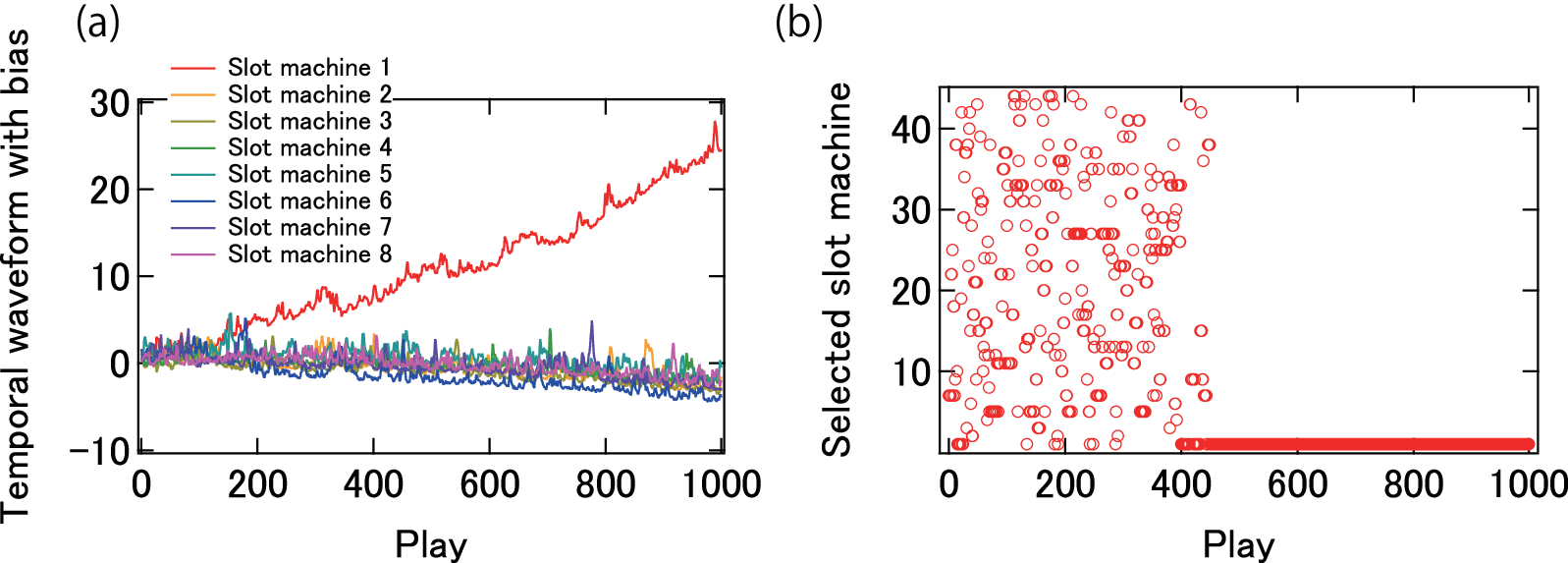}}
\caption{(a) An example of experimentally obtained temporal chaos with bias for $N_{\rm slot}$ = 8. (b) An example of selected slot machines for $N_{\rm slot}$ = 44. Sampling rate for the results is 2.5 GSa/s.}
\end{figure}

The CDR is determined while varying $N_{\rm slot}$. The CDR is computed 1000 times to minimize statistical variations. As depicted in Fig. 4(a), the CDR increases with $N_{\rm play}$, showing a sharp rise at specific points. As $N_{\rm slot}$ grows, the CDR decreases. To quantitatively assess the scaling to $N_{\rm slot}$, we examined $N_{\rm play}$ needed to reach a CDR of 0.95 thereafter, $N_{\rm CDR095}$, which is plotted as shown in Fig. 4(b). $N_{\rm CDR095}$ is 130, 190, 297, 519, 683 at $N_{\rm slot}$ of 4, 8, 16, 32, 44, respectively. Figure 4(b) also plots $N_{\rm CDR095}$ derived from software algorithms like Thompson sampling \cite{thompson1933likelihood} and UCB1-tuned \cite{auer2002finite}. The $N_{\rm CDR095}$ of the chaos comb is smaller than that of both Thompson sampling and UCB1-tuned, and it outperforms Thompson sampling by over 3 times at $N_{\rm CDR095}$ of 44. The regret at $N$th play, defined as \cite{iwami2022controlling}:
\begin{equation}
    \text{Regret}\left(N_{\rm play}\right) = N_{\rm play}P_{\rm max} - \sum_{m=1}^{N_{\rm slot}} P_m N_{\mathrm{play},m}
\end{equation}
is also calculated for both Thompson sampling and the chaos comb. In this equation, $P_{\rm max}$ and $P_m$ represent maximum hit probability and hit probability of $m$th machine, respectively. Regret indicates the missed rewards until $N_{\rm play}$. This regret is computed 1000 times to mitigate statistical variations. The regret at the 6000th play for varying slot machine counts is depicted in Fig. 4(c). For 44 slot machines, the chaos comb's regret is 2.6 times smaller than Thompson sampling's regret.

Another metric, the Shannon entropy at the $N$th play ($H\left(N_{\rm play}\right)$) is defined as \cite{iwami2022controlling}:
\begin{equation}
    H\left(N_{\rm play}\right) = -\sum_{m = 1}^{N_{\rm slot}} P_{\mathrm{sel,} m}\left(N_{\rm play}\right)\text{log}_2\left(P_{\mathrm{sel,} m}\left(N_{\rm play}\right)\right)
\end{equation}
\begin{equation}
    P_{\mathrm{sel,} m}\left(N_{\rm play}\right) = \frac{1}{W}\sum_{N = N_{\rm play} - W + 1}^{N_{\rm play}}\text{Sel}_m(N).
\end{equation}
Here, $W$ is the window size of $5N_{\rm slot}$, and $\text{Sel}_m(N)$ is a function that returns 1 if the $m$th slot machine is chosen at the $N$th play, and 0 otherwise. Shannon entropy, in essence, measures the variability in slot machine selections over a specified window. Figure 4(d) presents the Shannon entropy, averaged over 1000 iterations, when $N_{\rm slot}$ is 32. The chaos comb's entropy rapidly drops to an extremely low value, outpacing both UCB1-tuned and Thompson sampling. These results highlight the chaos comb's superiority over Thompson sampling, as it allows quicker exploration and yields more rewards.

\begin{figure}[!t]
\centering
\fbox{\includegraphics[width=0.95\linewidth]{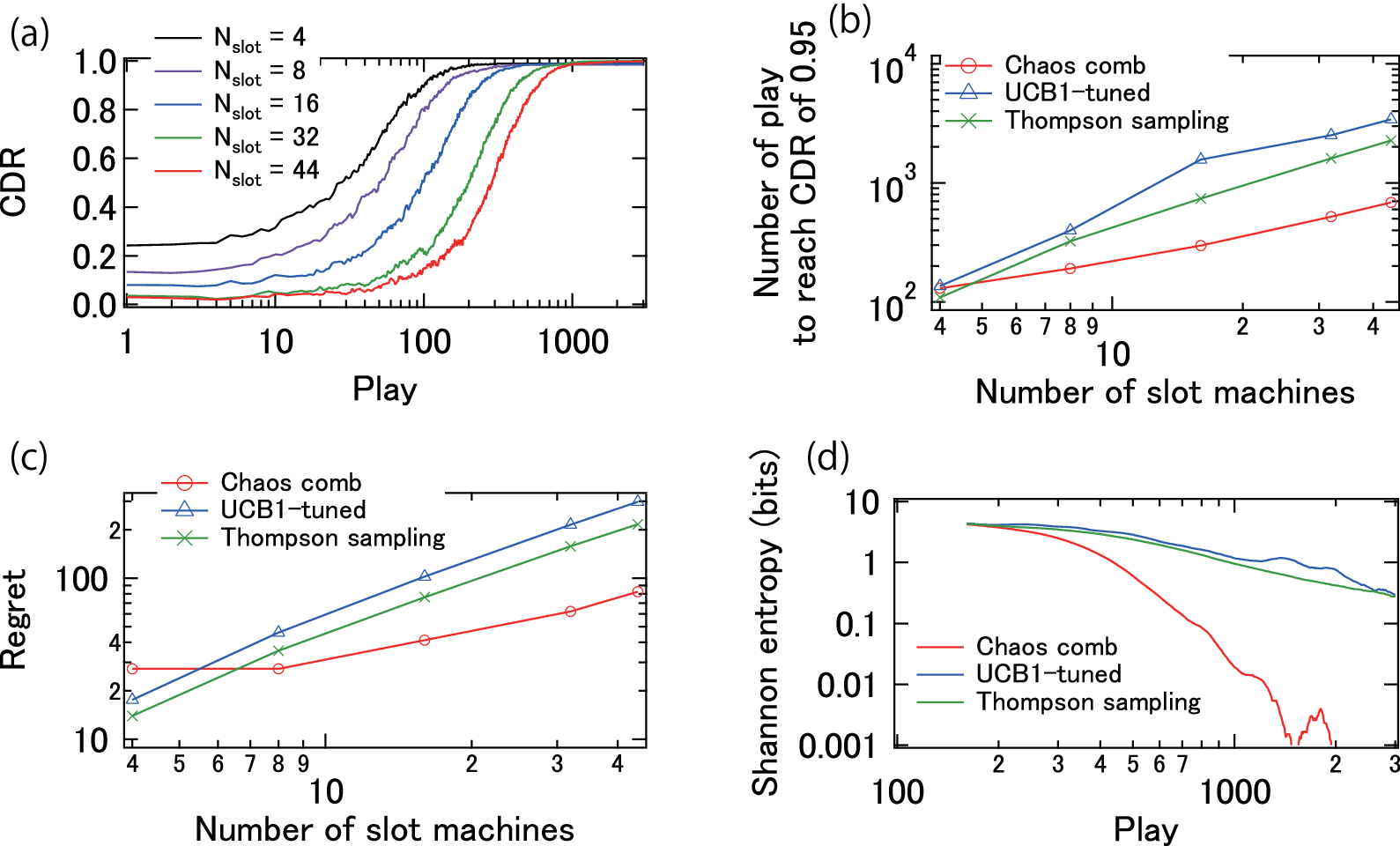}}
\caption{(a) Correct decision rate (CDR) for different $N_{\rm slot}$, derived from 44 comb modes of an experimentally obatained chaos comb. (b) $N_{\rm CDR095}$ of the chaos comb, UCB1-tuned and Thompson sampling. (c) Regret at of the 6000th play for different $N_{\rm slot}$. (d) Shannon entropy of $N_{\rm slot}$ = 32. The sampling rate and statistical averaging times for the results are 2.5 GSa/s and 1000, respectively.}
\end{figure}

\begin{figure}[!b]
\centering
\fbox{\includegraphics[width=0.95\linewidth]{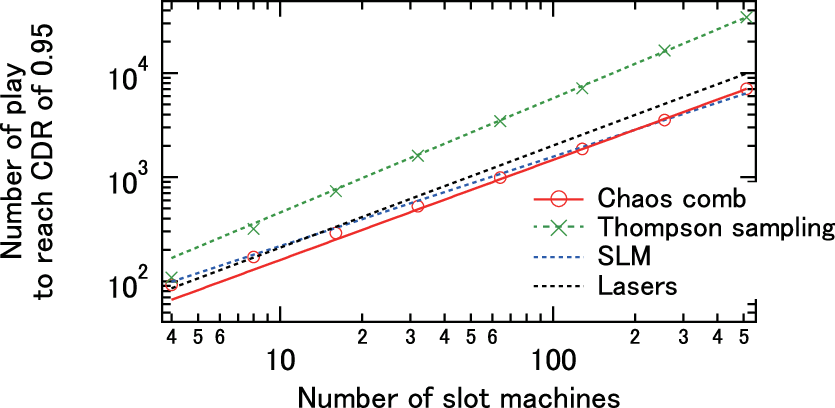}}
\caption{Comparison of $N_{\rm CDR095}$ across different methods. The red circles represent $N_{\rm CDR095}$ when multiple experimental temporal chaos from a chaos comb is collected from several comb modes in different time slots. The green 'x's are $N_{\rm CDR095}$ obtained from Thompson sampling. Fitted lines for the red circles and green 'x's are shown as the red solid line and green dashed line, respectively. The sampling rate for the red circles is 2.5 GSa/s. Both the red circles and the green 'x's use statistical averaging times of 1000. The blue line is $N_{\rm CDR095}$ ($= 30.0 \times N_{\rm slot}{}^{0.86}$) for the spatial-temporal chaos \cite{morijiri2023parallel}. The black line is $N_{\rm CDR095}$ ($= 22.0 \times N_{\rm slot}{}^{0.98}$) for the independent SIL lasers \cite{morijiri2022decision}.}
\end{figure}

To further examine the scaling to $N_{\rm slot}$, temporal chaos from several comb modes is experimentally obtained multiple times in different time slots. These are treated as independent temporal chaos and are assigned to individual slot machines. $N_{\rm CDR095}$ from the temporal chaos at $N_{\rm slot}$ of 4, 8, 16, 32, 64, 128, 256, and 512 is calculated, as shown in Fig. 5. $N_{\rm CDR095}$ offers a fitted curve of $17.4 \times N_{\rm slot}{}^{0.96}$. Compared to other photonic methods, such as using a multimode laser \cite{iwami2022controlling}, spatial temporal chaos in an SLM \cite{morijiri2023parallel}, and multiple self-injection locked semiconductor lasers \cite{morijiri2022decision}, the chaos comb displays the smallest 
$N_{\rm CDR095}$ up to roughly 300. For $N_{\rm play}$ greater than 300, the spatial temporal chaos in an SLM performs better due to its smaller exponent of 0.88.

\begin{figure}[!h]
\centering
\fbox{\includegraphics[width=0.7\linewidth]{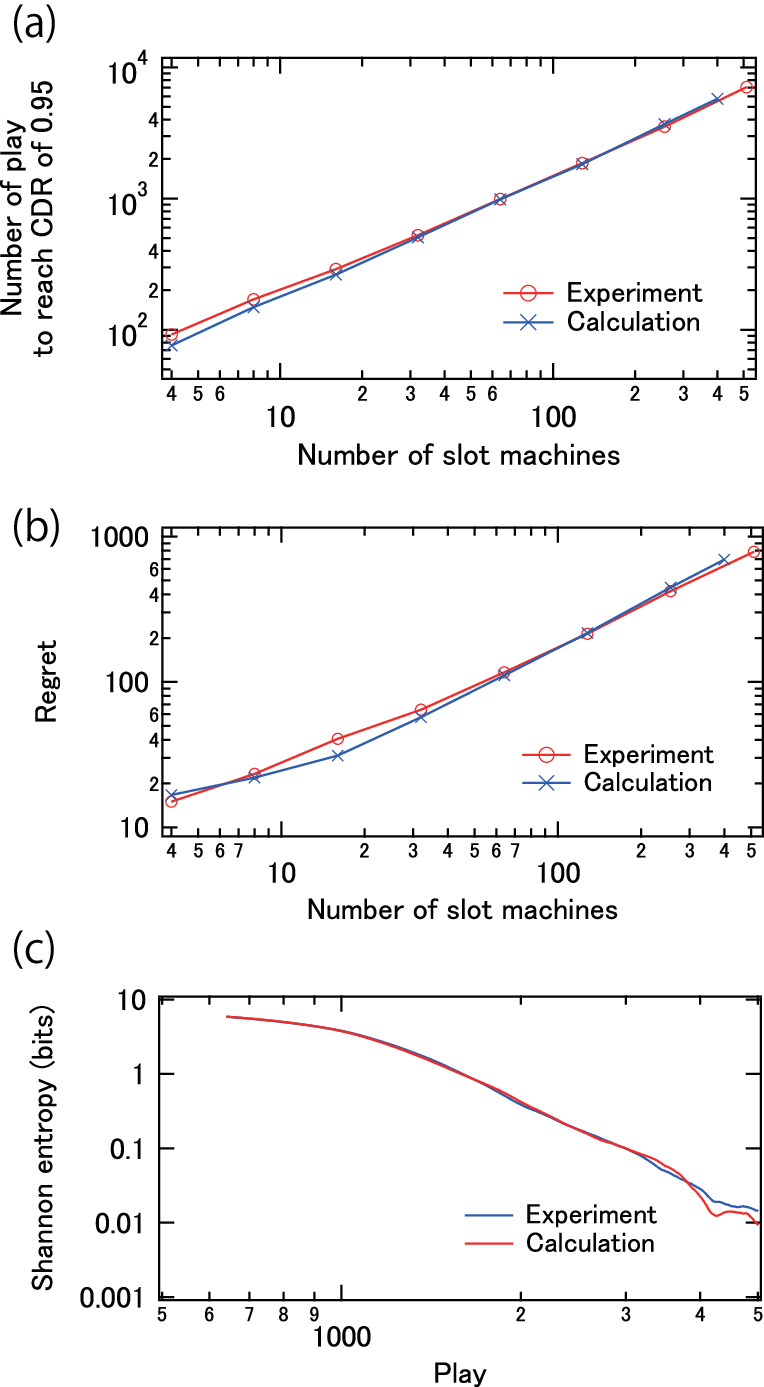}}
\caption{Comparison of MAB results between the experiment and (red curves) and numerical calculation (blue curves) (a) $N_{\rm CDR095}$, (b) regret at the 10000th play, and (c) Shannon entropy of $N_{\rm slot}$ = 128 . The experimental temporal chaos is the same as used in Fig. 5. The sampling rate and statistical averaging times both for the experiment and numerical calculation are 2.5 GSa/s and 1000, respectively.}
\end{figure}

Finally, numerical temporal chaos from a chaos comb, derived from 40 comb modes across 10 different time slots, is applied to the MAB. Given the well-matched characteristics of numerical temporal chaos with the experimental temporal chaos (as illustrated in Figs. 2(a) - (d)), metrics like $N_{\rm CDR095}$ (Fig. 6(a)), regret at the 10000th play (Fig. 6(b)), and Shannon entropy (Fig. 6(c)) closely resemble the experimental results. In most datasets, the discrepancy is within 10\%. The numerical calculation's $N_{\rm CDR095}$ offers a fitted curve of $16.1 \times N_{\rm slot}{}^{0.98}$, similar to the experimental result. It's worth noting that a large number of slot machines is necessary to accurately predict the scaling. Using a small number of slot machines might lead to an underestimation of the scaling, resulting in a smaller exponent. The fitted curves for the regret at the 10000th play are $2.65 \times N_{\rm slot}{}^{0.91}$ from the experimental result and $1.78 \times N_{\rm slot}{}^{1.0}$ from the numerical result. To estimate a more accurate exponent, a much larger number of slot machines is needed.

\section{Discussions}
The temporal chaos utilized in this paper exhibits a Poisson-like signal distribution, as depicted in Fig. 7(a). When temporal chaos is used for random bit generation \cite{Uchida2008}, symmetrical signal distribution is preferred \cite{Wang2017,Li2019}. To compare chaos with a Poisson-like distribution to chaos with a symmetrical distribution, the latter temporal chaos is created from the original chaos by subtracting the 2-bit delayed chaos from the original chaos (Fig. 7(b)). As shown in Fig. 7(c), there is no significant difference in the result of MAB. Unlike random-bit generation, symmetric signal distribution is not necessarily required, as used in the system of a multimode semiconductor laser \cite{iwami2022controlling} and spatial temporal chaos with an SLM and CMOS camera \cite{morijiri2023parallel}.

\begin{figure}[!h]
\centering
\fbox{\includegraphics[width=0.95\linewidth]{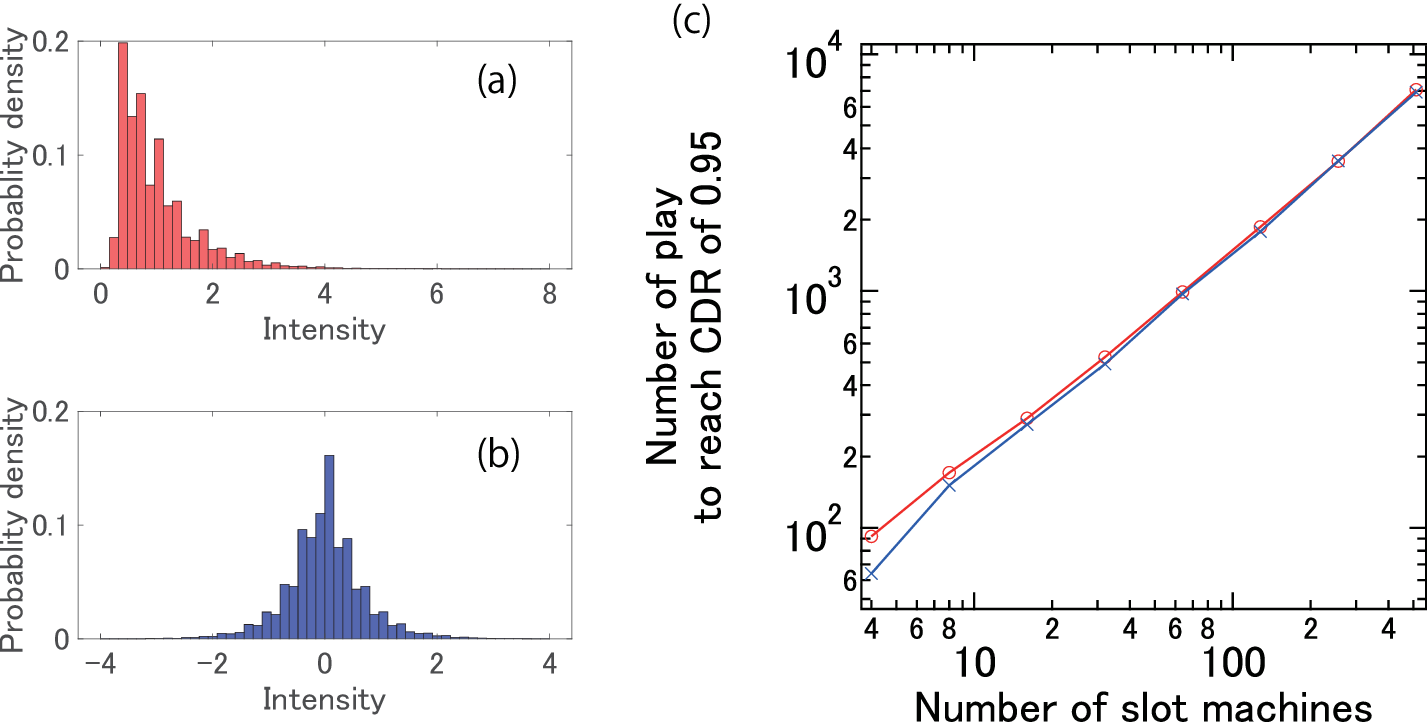}}
\caption{Histogram of the original, temporal chaos (a) and modified temporal chaos by subtracting 2-bit delayed temporal chaos from the original temporal chaos (b). (c) $N_{\rm CDR095}$ for the temporal chaos (red) and modified temporal chaos (blue). The original temporal chaos is experimentally obtained as used in Fig. 5. The sampling rate and statistical averaging times are 2.5 GSa/s and 1000, respectively.}
\end{figure}

The ultimate computation speed to reach a CDR of 0.95, i.e., ignoring offline-signal processing, is determined by $N_{\rm CDR095} \times \frac{1}{\text{Sampling rate}}$. Therefore, a faster sampling rate is better for the ultimate computation speed with a fixed $N_{\rm CDR095}$. Here, temporal chaos with different sampling rates is used to solve the MAB. Down-sampled temporal chaos with the sampling rates of 475 MSa/s and 2.375 GSa/s are created from a numerically calculated temporal chaos with a sampling rate of 9.5 GSa/s. $N_{\rm CDR095}$ from the three temporal chaos is calculated. As shown in Fig. 8, when $N_{\rm slot}$ is small (e.g., < 100), $N_{\rm CDR095}$ for 9.5 GSa/s is higher than that for 475 MSa/s and 2.375 GSa/s. The larger $N_{\rm CDR095}$ for the faster sampling rate is likely to the less randomness of the temporal chaos. However, when $N_{\rm slot}$ is large (e.g., > 100), $N_{\rm CDR095}$ from the three temporal chaos shows similar values.  This result indicates that having more comb modes using a smaller FSR at the cost of reduced RF spectrum bandwidth is a good strategy to increase $N_{\rm slot}$ from a chaos comb, as discussed next. 

\begin{figure}[!h]
\centering
\fbox{\includegraphics[width=0.8\linewidth]{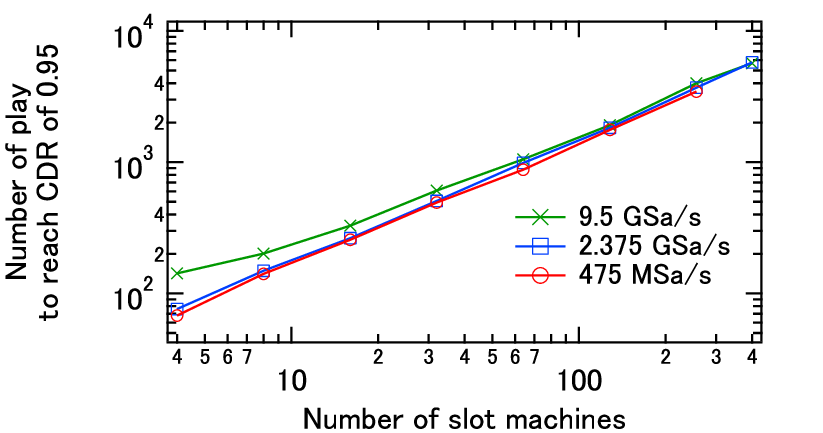}}
\caption{$N_{\rm CDR095}$ for the numerically-obtained temporal chaos with different sampling rates. Statistical averaging times is 1000.}
\end{figure}

To validate the aforementioned hypothesis, we numerically generated temporal chaos from chaos combs with distinct FSRs of 50 and 100 GHz. For an equitable comparison, both $F^2$ and $D_2$ were scaled proportionally to the FSR$^{-1}$ and FSR$^2$ , respectively \cite{Xuan2016,Herr2013}, under the assumption that the average pump power and waveguide dispersion remain constant. When the FSR is adjusted from 100 GHz to 50 GHz, the RF 10-dB bandwidth shrinks from 820 MHz to 540 MHz (Fig. 9(a)). Concurrently, the number of comb modes within the optical 10-dB bandwidth rises from 160 to 240 (Fig. 9(b)). This increase allows for the utilization of more comb modes from a chaos comb in addressing the MAB problem. While the comb mode power of the 50-GHz chaos comb is less than that of the 100-GHz chaos comb, the quantity of comb modes within the optical 10-dB bandwidth takes precedence. This is because comb mode power can be efficiently amplified using optical amplifiers. $N_{\rm CDR095}$ is computed with a consistent sampling rate of 2.5 GSa/s, and the results are presented in Fig. 9(c). The data reveals minimal variance between the 50-GHz and 100-GHz chaos combs. Notably, the $N_{\rm slot}$ for the 50-GHz chaos comb is 1.5 times that of the 100-GHz chaos comb. These findings bolster the proposition to reduce the FSR of microresonators in order to augment $N_{\rm slot}$, without adverse effects from the decline in RF bandwidth. While the study utilized only one side of the comb modes relative to the pump mode, employing both sides could effectively increase $N_{\rm slot}$. However, this might come with a potential rise in $N_{\rm CDR095}$ due to the positive correlation between the comb modes.

\begin{figure}[!h]
\centering
\fbox{\includegraphics[width=0.95\linewidth]{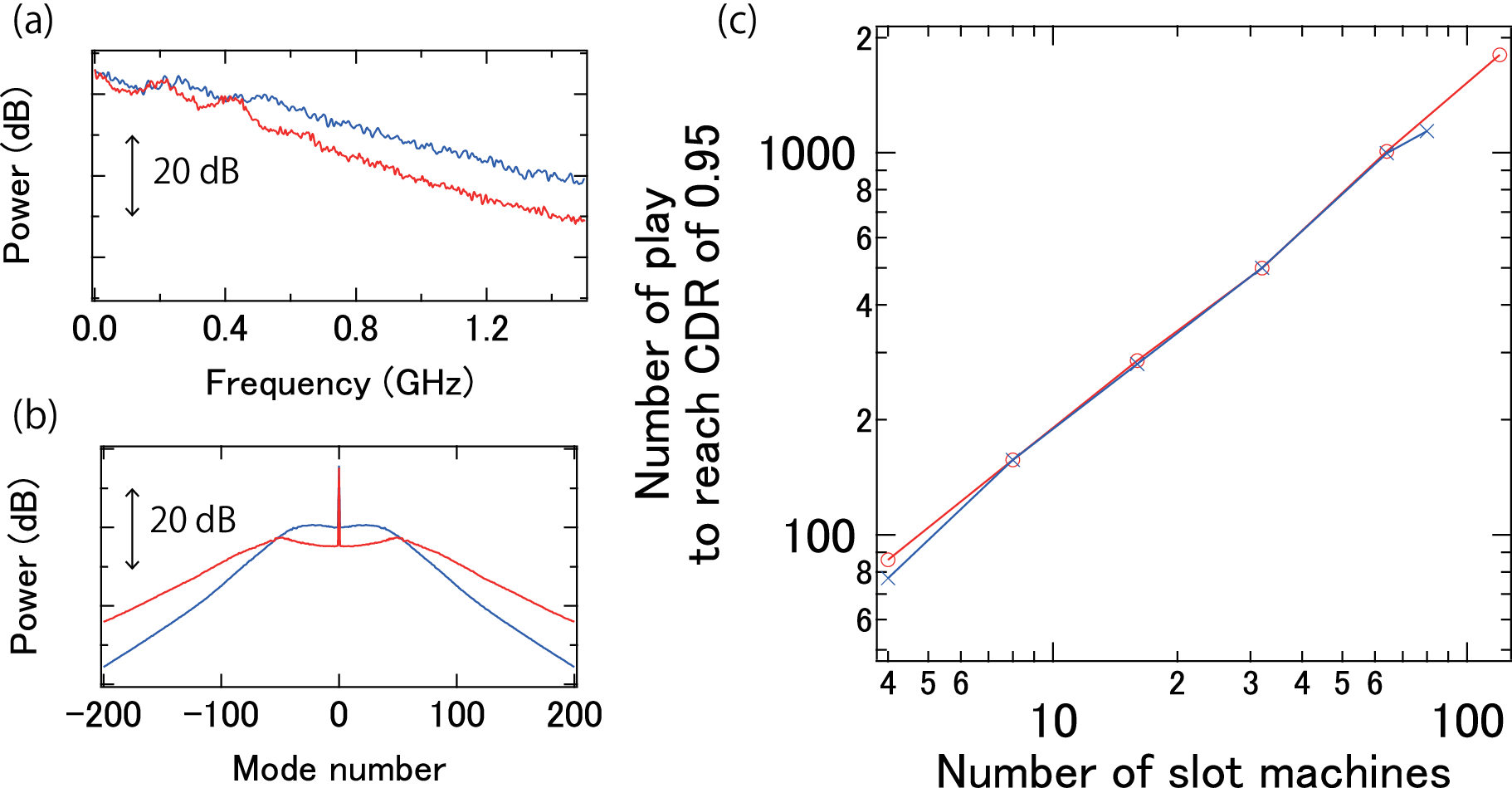}}
\caption{Comparison between chaos combs with FSRs of 50 GHz (red) and 100 GHz (blue). (a) RF spectra, (b) optical spectra, and (c) $N_{\rm CDR095}$ for the numerically-obtained chaos combs. The sampling rate and statistical averaging times for $N_{\rm CDR095}$ are 2.5 GSa/s and 1000, respectively.}
\end{figure}

Another approach to augment $N_{\rm slot}$ involves the utilization of multiple chaos combs. This strategy doesn't increase system complexity, especially in terms of chaos comb generation, given the ease with which multiple microresonators can be fabricated on a single chip. Only one pump CW laser might be necessary. This is because there's an absence of coherence between the comb modes in different chaos combs, as well as among the comb modes within the same chaos comb.

The temporal response of our temporal chaos is constrained by the Si$_3$N$_4$ microresonator. By employing microresonators with a higher nonlinear index, such as AlGaAs \cite{Chang2020}, a temporal response that is 10 times faster can be achieved with the same FSR and pump power as used in our experiments. This would enable the use of a microresonator with an FSR of a few tens of GHz. Such chaos combs could produce several hundred comb modes within a 10-dB optical bandwidth, scaling $N_{\rm slot}$ to levels unparalleled by other photonic methods.

Decision making using a chaos comb has been very recently reported by Shen et al. \cite{Shen2023}. Their study, however, primarily focused on random number generation and CDR evaluation for decision making without a thorough analysis of the properties crucial for accurate decision making. In contrast, we offered a detailed analysis of decision making, including aspects such as CDR, regret, and entropy. Moreover, chaos combs were derived both experimentally and numerically, showing strong correspondences in the characteristic of the chaos combs obtained both ways, as well as in the results of MAB derived from these chaos combs. This validation underpins discussions about the sampling rate, and the influence of RF/optical bandwidth on solving MAB. Our findings of such insights are essential for optimizing the parameters of chaos combs and measurements for MAB, with the goals of increasing $N_{\rm slot}$ and minimizing the ultimate computation time.

\section{Conclusions}
We have demonstrated an approach to tackle the MAB problem using a chaotic microresonator comb. The uniqueness of our method lies in the use of temporal chaos from different comb modes of a chaos comb, enabling the simultaneous handling of multiple slot machines. Both for experimentally-obtained and numerically-calculated chaos comb, our system outperforms conventional software algorithms, and it also showcases superior results compared to other photonic approaches. The potential scalability of our method, combined with the inherent speed of photonics, points to a promising avenue for real-time, high-throughput decision-making applications. Future work will focus on integrating the entire system on a photonic integrated circuit, further enhancing its compactness and robustness.

\section*{acknowledgments}
We would like to acknowledge Kazutaka Kanno for fruitful discussions. We acknowledge financial supports from Japan Society for the Promotion of Science (23H04806) and Cabinet Office, Government of Japan (Subsidy for Reg. Univ. and Reg. Ind. Creation).

\ifCLASSOPTIONcaptionsoff
  \newpage
\fi



%



\bibliography{scibib}

\bibliographystyle{osajnl}

%








\end{document}